\documentclass[twocolumn,showpacs,prb]{revtex4}%
\usepackage{amsfonts}
\usepackage{amsmath}
\usepackage{amssymb}
\usepackage{graphicx}
\usepackage{times}
\usepackage[colorlinks]{hyperref}
\providecommand{\U}[1]{\protect\rule{.1in}{.1in}}

\begin{document}
\title{Current-induced spin polarization in spin-orbit-coupled electron systems}
\author{Ming-Hao Liu}%
\thanks{Present address: No. 2-1, Fushou Lane, Chengsiang Village, Gangshan Township,
Kaohsiung County 82064, Taiwan}%
\email{d92222010@ntu.edu.tw}%
\affiliation{Department of Physics,
National Taiwan University, Taipei 10617, Taiwan}
\author{Son-Hsien Chen}
\affiliation{Department of Physics, National Taiwan University,
Taipei 10617, Taiwan}
\author{Ching-Ray Chang}
\affiliation{Department of Physics, National Taiwan University,
Taipei 10617, Taiwan}
\begin{abstract}
Current-induced spin polarization (CISP) is rederived in ballistic
spin-orbit-coupled electron systems, based on equilibrium
statistical mechanics. A simple and useful picture is
correspondingly proposed to help understand the CISP and predict the
polarization direction. Nonequilibrium Landauer-Keldysh formalism is
applied to demonstrate the validity of the statistical picture,
taking the linear Rashba-Dresselhaus [001] two-dimensional system as
a specific example. Spin densities induced by the CISP in
semiconductor heterostructures and in metallic surface states are
compared, showing that the CISP increases with the spin splitting
strength and hence suggesting that the CISP should be more
observable on metal and semimetal surfaces due to the discovered
strong Rashba splitting. An application of the CISP designed to
generate a spin-Hall pattern in the inplane, instead of the
out-of-plane, component is also proposed.
\end{abstract}
\volumeyear{year}
\volumenumber{number}
\issuenumber{number}
\eid{identifier}
\date{\today}
\maketitle

\section{Introduction}

The aim of preparing and controlling spins in all-electrical nonmagnetic
devices has been shown to be possible in semiconducting bulk and
two-dimensional electron systems (2DESs).\cite{Awschalom02,Kato04a} Besides
the optical spin injection, a much more natural way of spin orientation is to
make use of the spin-orbit (SO) coupling due to the lack of inversion symmetry
of the underlying material.\cite{Winkler03} When passing an unpolarized
electric current (electrons carrying random spins) through an SO-coupled
material, spin-dependent consequences arise, among which two famous phenomena
are the spin-Hall effect
(SHE)\cite{Dyakonov71a,Hirsch99,Murakami03,Sinova04,Kato04,Wunderlich05} and
the current-induced spin polarization (CISP).

In the CISP phenomenon, unpolarized electric current is expected to be
spin-polarized when flowing in a SO-coupled sample. This effect was first
theoretically proposed in the early 90s. Edelstein\cite{Edelstein90} employed
linear-response theory to calculate the spin polarization due to an electric
current in the presence of SO coupling linear in momentum, taking into account
low-concentration impurities. Aronov and Lyanda-Geller\cite{Aronov89} solved
the quantum Liouville's theorem for the spin density matrix to show the CISP,
taking into account scattering as well. Recently, the CISP phenomenon has been
experimentally proven.\cite{CISPexp1,CISPexp2,CISPexp3} Moreover, both the SHE
and CISP have been observed at room temperature.\cite{CISPexp4}

\begin{figure}[b]
\centering\includegraphics[width=4.5cm]{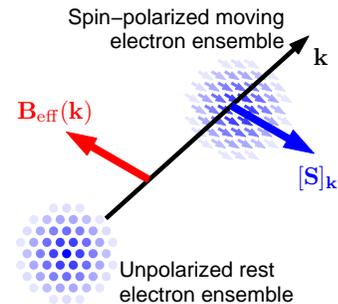} \caption{{}(Color online)
Statistical picture of the current-induced spin polarization phenomenon.}%
\label{fig1}%
\end{figure}

In this paper we propose another viewpoint based on equilibrium statistical
mechanics to explain the CISP in the absence of impurity scattering, for both
bulk and two-dimensional systems. We show that the canonical ensemble average
(CEA) of electrons moving with a wave vector $\mathbf{k}$ immediately
prescribes a spin polarization antiparallel to the effective magnetic field
$\mathbf{B}_{\text{eff}}(\mathbf{k})$ stemming from the underlying SO coupling
not necessarily linear in $k$, and hence explains the CISP. Correspondingly, a
much simpler picture, compared to the early theoretical works of Refs.
\onlinecite{Edelstein90}
and
\onlinecite{Aronov89}%
, helps provide a qualitative and straightforward explanation for the CISP: In
an SO coupled 2DES without external magnetic field, an ensemble of rest
electrons is unpolarized, while it becomes spin-polarized antiparallel to
$\mathbf{B}_{\text{eff}}(\mathbf{k})$ when moving along $\mathbf{k}$ (see Fig.
\ref{fig1}).

To demonstrate the validness of this elementary statistical argument, spin and
charge transports in finite-size four-terminal conducting 2DESs with Rashba
and linear Dresselhaus [001] SO couplings, are numerically analyzed using the
more sophisticated Landauer-Keldysh formalism
(LKF),\cite{Datta95,Nikolic05,Nikolic06} allowing for nonequilibrium
statistics. Good agreement between the analytical CEA and the numerical LKF
will be seen, consolidating our statistical picture. In addition to the
semiconducting heterostructures, we also extend the analysis of the CISP to
metal and semimetal surfaces, and compare the polarization strengths. Finally,
an application of the CISP, resembling an inplane SHE, will be subsequently
proposed. Throughout this paper, all the band parameters used in the LKF are
extracted from experiments by matching the band structures calculated by the
tight-binding model (and hence the density of states calculated by the LKF)
with the experimentally measured ones.\cite{Liu07a}

This paper is organized as follows. In Sec. \ref{sec analytical}, we discuss
the general properties of the system with SO coupling and derive the CISP in
the ballistic limit using statistical mechanics. In Sec. \ref{sec numerical}
the LKF is applied partly to examine the validity of the statistical picture
of the CISP introduced in Sec. \ref{sec analytical}, and partly for further
investigation. Summary of the present work will be given in Sec.
\ref{sec summary}.

\section{Analytical derivations\label{sec analytical}}

Consider a SO-coupled system, subject to the single-particle Hamiltonian%
\begin{equation}
\mathcal{H}=\frac{\hbar^{2}\mathbf{k}^{2}}{2m}\openone+\mathbf{S}\cdot
\vec{\Omega}\left(  \mathbf{k}\right)  , \label{H}%
\end{equation}
where $m$ is the effective mass, $\openone$ is the $2\times2$ identity matrix,
$\mathbf{S}=(\hbar/2)\vec{\sigma}$ is the spin operator, $\vec{\sigma}%
\equiv(\sigma^{x},\sigma^{y},\sigma^{z})$ being the Pauli matrix vector, and
$\vec{\Omega}\left(  \mathbf{k}\right)  =(e/mc)\mathbf{B}_{\text{eff}}\left(
\mathbf{k}\right)  $ is the momentum-dependent Larmor frequency vector, with
$\mathbf{B}_{\text{eff}}\left(  \mathbf{k}\right)  $ being the effective
magnetic field stemming from the SO coupling.\cite{Zutic04}

\subsection{Larmor frequency vectors}

For III-V (zinc blende) bulk semiconductors,\cite{Dresselhaus55} the Larmor
frequency in Eq. (\ref{H}) is written as\cite{Dyakonov71}%
\begin{equation}
\vec{\Omega}\left(  \mathbf{k}\right)  =\frac{\eta\hbar^{2}}{\left(
2m^{3}E_{g}\right)  ^{1/2}}\vec{\kappa}, \label{Larmor}%
\end{equation}
where $\eta$ is a dimensionless parameter specifying the spin-orbit coupling
strength, $E_{g}$ is the band gap, and $\vec{\kappa}$ is given by%
\begin{equation}
\vec{\kappa}=\left(
\begin{array}
[c]{c}%
k_{x}(k_{y}^{2}-k_{z}^{2})\\
k_{y}(k_{z}^{2}-k_{x}^{2})\\
k_{z}(k_{x}^{2}-k_{y}^{2})
\end{array}
\right)  . \label{kappa}%
\end{equation}
Here $k_{i}$'s are the wave vector components along the crystal principle axes.

When restricted to two-dimension, the component of the wave vector normal to
the 2DES is averaged. For [001] quantum wells, one has $k_{z}^{2}%
\rightarrow\langle k_{z}^{2}\rangle$ and $k_{z}\rightarrow\langle k_{z}%
\rangle=\langle i\partial_{z}\rangle=0$ to rewrite Eq. (\ref{kappa}) as
$\vec{\kappa}^{[001]}=[k_{x}(k_{y}^{2}-\langle k_{z}^{2}\rangle),k_{y}(\langle
k_{z}^{2}\rangle-k_{x}^{2}),0]$, so that the Larmor frequency (\ref{Larmor})
takes the form%
\begin{equation}
\vec{\Omega}^{[001]}=\frac{2\beta}{\hbar}\left(  -k_{x},k_{y},0\right)
+\frac{2\beta}{\hbar\langle k_{z}^{2}\rangle}\left(  k_{x}k_{y}^{2}%
,-k_{y}k_{x}^{2},0\right)  , \label{Larmor [001]}%
\end{equation}
where $\beta$ is defined by%
\begin{equation}
\beta=\frac{\hbar}{2}\frac{\eta\hbar^{2}}{\left(  2m^{3}E_{g}\right)  ^{1/2}%
}\langle k_{z}^{2}\rangle=\gamma\langle k_{z}^{2}\rangle\label{beta}%
\end{equation}
and is referred to as the Dresselhaus SO coupling constant. The $\gamma$
parameter (corresponding to $\mathfrak{b}_{41}^{6c6c}$ of Ref.
\onlinecite{Winkler03}) is material-dependent and is roughly $27$ $%
\operatorname{eV}%
\operatorname{\text{\AA}}%
^{3}$ for both GaAs and InAs.\cite{Knap96,Winkler03}

The first term in Eq. (\ref{Larmor [001]}),%
\begin{equation}
\vec{\Omega}_{D}^{[001]}=\frac{2\beta}{\hbar}\left(  -k_{x},k_{y},0\right)  ,
\label{Larmor D}%
\end{equation}
is the linear Dresselhaus [001] term, which will dominate for small $k$
region. The corresponding SO term $\mathcal{H}_{D}^{[001]}=\mathbf{S}\cdot
\vec{\Omega}_{D}^{[001]}=\beta(-k_{x}\sigma^{x}+k_{y}\sigma^{y})$ is known as
the linear Dresselhaus [001] model Hamiltonian.\cite{Winkler03,Zutic04} With
larger $k$ the second term in Eq. (\ref{Larmor [001]})---the $k^{3}$
term---becomes important. We will come back to this later. For other quantum
wells such as [110] and [111], the $\vec{\kappa}$ vector given by Eq.
(\ref{kappa}) can be recast into a form that depends on the growth direction
$\hat{n}$ of the 2DES.\cite{Dyakonov86} (See also Ref. \onlinecite{Zutic04}.)

When writing the Larmor frequency vector as%
\begin{equation}
\vec{\Omega}_{R}=\frac{2\alpha}{\hbar}(\mathbf{k}\times\hat{n}),
\label{Larmor R}%
\end{equation}
the linear Rashba model Hamiltonian\cite{Winkler03,Zutic04,Bychkov84}
$\mathcal{H}_{R}=\mathbf{S}\cdot\vec{\Omega}_{R}=\alpha(\mathbf{k}\times
\hat{n})$ is recovered. Here $\alpha$ is the Rashba SO coupling constant.

\subsection{Time-reversal symmetry}

Before deriving the CISP, we provide the following two intrinsic properties of
the Hamiltonian (\ref{H}). First, we show that the contribution to the SO
terms in solid is odd in $k$ due to time-reversal symmetry, which is also
remarked in Ref. \onlinecite{Winkler03}. For spin-1/2 systems subject to
Hamiltonian (\ref{H}), the energy dispersion can be written as%
\begin{equation}
E_{\sigma}(\mathbf{k})=E_{0}+\sigma\Delta_{\mathbf{k}}, \label{E}%
\end{equation}
where $E_{0}=\hbar^{2}k^{2}/2m$ is the kinetic energy, $\sigma=\pm1$ is the
spin state label, and $\Delta_{\mathbf{k}}$ is the spin splitting due to SO
coupling. In the absence of external magnetic field, the time-reversal
symmetry is preserved, resulting in $E_{+}(\mathbf{k})=E_{-}(-\mathbf{k})$,
or,%
\begin{equation}
+\Delta_{\mathbf{k}}=-\Delta_{-\mathbf{k}}, \label{TimeReversal}%
\end{equation}
which implies that nonvanishing spin splitting $\Delta_{\mathbf{k}}$ is odd in
$k$. Note that Eq. (\ref{TimeReversal}) also implies%
\begin{equation}
\vec{\Omega}\left(  -\mathbf{k}\right)  =-\vec{\Omega}\left(  \mathbf{k}%
\right)  , \label{O(-k) = -O(k)}%
\end{equation}
which agrees with our intuition. Apparently, Eq. (\ref{O(-k) = -O(k)}) is
obeyed by all the previously reviewed Larmor frequency vectors.

Second, we show $\langle\pm,\mathbf{k}|\vec{\sigma}|\pm,\mathbf{k}%
\rangle=-\langle\mp,\mathbf{k}|\vec{\sigma}|\mp,\mathbf{k}\rangle$, where
$|\sigma,\mathbf{k}\rangle$ is the eigenstate of Hamiltonian (\ref{H}). We
begin with the Schr\"{o}dinger equation,%
\begin{equation}
\mathcal{H}|\sigma,\mathbf{k}\rangle=(\frac{\hbar^{2}\mathbf{k}^{2}}%
{2m}\openone+\mathbf{S}\cdot\vec{\Omega}\left(  \mathbf{k}\right)
)|\sigma,\mathbf{k}\rangle=E_{\sigma}(\mathbf{k})|\sigma,\mathbf{k}\rangle.
\label{SE}%
\end{equation}
Comparing Eq. (\ref{SE}) with Eq. (\ref{E}), we deduce $\mathbf{S}\cdot
\vec{\Omega}\left(  \mathbf{k}\right)  |\sigma,\mathbf{k}\rangle=\sigma
\Delta_{\mathbf{k}}|\sigma,\mathbf{k}\rangle$, or,%
\begin{equation}
\langle\sigma,\mathbf{k}|\mathbf{S}\cdot\vec{\Omega}\left(  \mathbf{k}\right)
|\sigma,\mathbf{k}\rangle=\sigma\Delta_{\mathbf{k}}, \label{<SdotO>}%
\end{equation}
where $|\sigma,\mathbf{k}\rangle$ is assumed normalized. This implies%
\begin{equation}
\langle+,\mathbf{k}|\mathbf{S}\cdot\vec{\Omega}\left(  \mathbf{k}\right)
|+,\mathbf{k}\rangle=-\langle-,\mathbf{k}|\mathbf{S}\cdot\vec{\Omega}\left(
\mathbf{k}\right)  |-,\mathbf{k}\rangle. \label{<+,k> = -<-,k> 2}%
\end{equation}
Factoring out and canceling $\vec{\Omega}(\mathbf{k})$ on both sides, we
arrive at%
\begin{equation}
\langle+,\mathbf{k}|\vec{\sigma}|+,\mathbf{k}\rangle=-\langle-,\mathbf{k}%
|\vec{\sigma}|-,\mathbf{k}\rangle. \label{<+,k> = -<-,k>}%
\end{equation}
Equation (\ref{<+,k> = -<-,k>}) is a general property of Eq. (\ref{H}) and is
valid for systems with dispersions $E_{\sigma}\left(  \mathbf{k}\right)
=E_{0}+\sigma\Delta_{\mathbf{k}}$, where the spin splitting $\Delta
_{\mathbf{k}}$ is not necessarily linear in $k$. This property
(\ref{<+,k> = -<-,k>}) will play a tricky role in the coming derivation of the
CISP based on statistical mechanics in Sec. \ref{sec CEA}.

Note that Eq. (\ref{<+,k> = -<-,k>}) is also a consequence of time-reversal
symmetry (\ref{TimeReversal}), as one can easily prove as follows. Using Eq.
(\ref{<SdotO>}) we rewrite Eq. (\ref{TimeReversal}) as%
\begin{equation}
\langle+,\mathbf{k}|\mathbf{S}\cdot\vec{\Omega}\left(  \mathbf{k}\right)
|+,\mathbf{k}\rangle=\langle-,-\mathbf{k}|\mathbf{S}\cdot\vec{\Omega}\left(
-\mathbf{k}\right)  |-,-\mathbf{k}\rangle. \label{TimeReversal 2}%
\end{equation}
Equation (\ref{<SdotO>}) also implies%
\begin{equation}
\langle\sigma,\mathbf{k}|\mathbf{S}\cdot\vec{\Omega}\left(  \mathbf{k}\right)
|\sigma,\mathbf{k}\rangle=-\langle-\sigma,\mathbf{k}|\mathbf{S}\cdot
\vec{\Omega}\left(  \mathbf{k}\right)  |-\sigma,\mathbf{k}\rangle
\label{for proof 1}%
\end{equation}
when one regards $\sigma\Delta_{\mathbf{k}}$ as $-\left(  -\sigma\right)
\Delta_{\mathbf{k}}$. In addition, Eq. (\ref{TimeReversal}) implies%
\begin{equation}
\langle\sigma,\mathbf{k}|\mathbf{S}\cdot\vec{\Omega}\left(  \mathbf{k}\right)
|\sigma,\mathbf{k}\rangle=\langle\sigma,-\mathbf{k}|\mathbf{S}\cdot\vec
{\Omega}\left(  \mathbf{k}\right)  |\sigma,-\mathbf{k}\rangle
\label{for proof 2}%
\end{equation}
because of%
\begin{subequations}
\begin{align}
\langle\sigma,\mathbf{k}|\mathbf{S}\cdot\vec{\Omega}\left(  \mathbf{k}\right)
|\sigma,\mathbf{k}\rangle &  =\sigma\Delta_{\mathbf{k}}=-\sigma\Delta
_{-\mathbf{k}}\nonumber\\
&  =\langle-\sigma,-\mathbf{k}|\mathbf{S}\cdot\vec{\Omega}\left(
-\mathbf{k}\right)  |-\sigma,-\mathbf{k}\rangle\nonumber\\
&  =-\langle\sigma,-\mathbf{k}|\mathbf{S}\cdot\vec{\Omega}\left(
-\mathbf{k}\right)  |\sigma,-\mathbf{k}\rangle\label{for proof 3}\\
&  =\langle\sigma,-\mathbf{k}|\mathbf{S}\cdot\vec{\Omega}\left(
\mathbf{k}\right)  |\sigma,-\mathbf{k}\rangle, \label{for proof 4}%
\end{align}
where Eqs. (\ref{for proof 1}) and (\ref{O(-k) = -O(k)}) are used in
(\ref{for proof 3}) and (\ref{for proof 4}), respectively. Substituting Eqs.
(\ref{O(-k) = -O(k)}) and (\ref{for proof 2}) into Eq. (\ref{TimeReversal 2}),
we obtain Eq. (\ref{<+,k> = -<-,k> 2}), and hence the property
(\ref{<+,k> = -<-,k>}).

\subsection{Current-induced spin polarization by canonical ensemble average}

Having seen the general properties of the Hamiltonian (\ref{H}) under the
time-reversal symmetry, we now derive the equilibrium statistics version of
the CISP. In quantum statistics, any physical quantity, say $A$, is expressed
in terms of the quantum statistical average $\left[  A\right]
=\operatorname{Tr}(\rho A).$ Adopting the canonical ensemble, the average
reads
\end{subequations}
\begin{equation}
\left[  A\right]  =\frac{\operatorname{Tr}\left(  e^{-\mathcal{H}/k_{B}%
T}A\right)  }{\sum_{\nu}e^{-E_{\nu}/k_{B}T}}, \label{EA}%
\end{equation}
where $k_{B}$ is the Boltzmann constant, $T$ is temperature, $\nu$ is a
quantum number labeling the states, and $E_{\nu}$ is the eigenenergy of state
$\nu$ solved from Hamiltonian $\mathcal{H}$.

Now consider an unpolarized electron ensemble in a 2DES, subject to
Hamiltonian (\ref{H}). Our main interest here is the CEA of the spin operators
of an ensemble of electrons, subject to an identical wave vector $\mathbf{k}$.
By this we mean that the summation in Eq. (\ref{EA}) runs over the spin index
$\sigma$ only. This gives
\[
\lbrack\mathbf{S}]_{\mathbf{k}}=\frac{\hbar}{2}\frac{\operatorname{Tr}%
(e^{-\mathcal{H}/k_{B}T}\vec{\sigma})}{\sum_{\sigma=\pm}e^{-E_{\sigma}\left(
\mathbf{k}\right)  /k_{B}T}}.
\]
Choosing the basis $|\sigma,\mathbf{k}\rangle$ for the trace, one is led to
\[
\lbrack\mathbf{S]}_{\mathbf{k}}=\frac{\hbar}{2}\frac{\sum_{\sigma
}e^{-E_{\sigma}\left(  \mathbf{k}\right)  /k_{B}T}\langle\sigma,\mathbf{k}%
|\vec{\sigma}|\sigma,\mathbf{k}\rangle}{\sum_{\sigma}e^{-E_{\sigma}\left(
\mathbf{k}\right)  /k_{B}T}}.
\]
Using the property (\ref{<+,k> = -<-,k>}) and factoring out $e^{-\hbar
^{2}k^{2}/2mk_{B}T}$ from $e^{-E_{\sigma}/k_{B}T},$ we arrive at the general
expression%
\begin{equation}
\lbrack\mathbf{S}]_{\mathbf{k}}=-\frac{\hbar}{2}\tanh\frac{\Delta_{\mathbf{k}%
}}{k_{B}T}\langle+,\mathbf{k}|\vec{\sigma}|+,\mathbf{k}\rangle. \label{Sk}%
\end{equation}
To re-express Eq. (\ref{Sk}) in terms of the effective magnetic field
$\mathbf{B}_{\text{eff}}(\mathbf{k})$, defined by%
\begin{equation}
\mathbf{B}_{\text{eff}}(\mathbf{k})\equiv\frac{mc}{e}\Omega(\mathbf{k}%
)=\frac{\dfrac{\hbar}{2}\Omega(\mathbf{k})}{\mu_{B}}, \label{Beff}%
\end{equation}
we rewrite Eq. (\ref{<SdotO>}) with $\sigma=+1$ as%
\begin{equation}
\langle+,\mathbf{k}|\vec{\sigma}|+,\mathbf{k}\rangle\cdot\mathbf{B}%
_{\text{eff}}(\mathbf{k})=\frac{\Delta_{\mathbf{k}}}{\mu_{B}}.
\label{<+,k> dot Beff}%
\end{equation}
Noting $|\langle\sigma,\mathbf{k}|\vec{\sigma}|\sigma,\mathbf{k}\rangle|=1$
(unit vector) and $\left\vert (\hbar/2)\Omega(\mathbf{k})\right\vert
=\Delta_{\mathbf{k}}$, Eq. (\ref{<+,k> dot Beff}) implies%
\begin{equation}
\langle+,\mathbf{k}|\vec{\sigma}|+,\mathbf{k}\rangle=\mathbf{\hat{B}%
}_{\text{eff}}(\mathbf{k}), \label{<+,k> = Beff hat}%
\end{equation}
i.e., the direction of the effective magnetic field. Therefore, Eq. (\ref{Sk})
can be written as%
\begin{equation}
\lbrack\mathbf{S}]_{\mathbf{k}}=-\frac{\hbar}{2}\tanh\frac{\Delta_{\mathbf{k}%
}}{k_{B}T}\mathbf{\hat{B}}_{\text{eff}}(\mathbf{k}), \label{Sk 2}%
\end{equation}
which is exactly the analog of the CEA of electron spin in vacuum subject to
an applied magnetic field.\cite{Sakurai94}

Equation (\ref{Sk 2}) now has a transparent meaning: In the presence of SO
coupling, an ensemble of rest electrons ($\mathbf{k}\rightarrow0$) is
unpolarized since $\Delta_{\mathbf{k}\rightarrow0}=0$, while it becomes
spin-polarized antiparallel to $\mathbf{\hat{B}}_{\text{eff}}(\mathbf{k})$
when moving along $\mathbf{k}$. This picture is schematically shown in Fig.
\ref{fig1}. Moreover, the hyperbolic tangent factor $\tanh(\Delta_{\mathbf{k}%
}/k_{B}T)$ clearly predicts the decrease with $T$ and the increase with
$\Delta_{\mathbf{k}}$ in the polarization magnitude, and therefore explains
two signatures of the CISP qualitatively: (i) The CISP may persist up to the
room temperature. Taking $\Delta_{\mathbf{k}}\approx3.68$ m$%
\operatorname{eV}%
$ from Ref. \onlinecite{CISPexp3}, one has $\tanh[\Delta_{\mathbf{k}}%
/(k_{B}\times300%
\operatorname{K}%
)]/\tanh[\Delta_{\mathbf{k}}/(k_{B}\times10%
\operatorname{K}%
)]\approx14\%$. (ii) As $\langle\mathbf{k}\rangle\propto V_{0}$ (Ref.
\onlinecite{CISPexp2}%
) implies $\Delta_{\mathbf{k}}\propto V_{0}$, the magnitude of the CISP
governed by $\tanh(\Delta_{\mathbf{k}}/k_{B}T)$ is supposed to increase with
the bias, as is experimentally proven.\cite{CISPexp1}

\subsection{Explicit forms of current-induced spin polarization}

From Eq. (\ref{Sk 2}), it is now clear that the direction of the CISP is given
by the effective magnetic field direction $\mathbf{\hat{B}}_{\text{eff}%
}(\mathbf{k})$. Alternatively, one can use the direction of the Larmor
frequency vector, $\hat{\Omega}(\mathbf{k})$, to describe the CISP direction
since $\mathbf{B}_{\text{eff}}(\mathbf{k})$ and $\vec{\Omega}(\mathbf{k})$
are, by definition of Eq. (\ref{Beff}), collinear. Therefore, the CISP
direction in III-V bulk semiconductors is given by Eq. (\ref{kappa}%
).\begin{figure}[t]
\centering
\includegraphics[width=8.65cm]{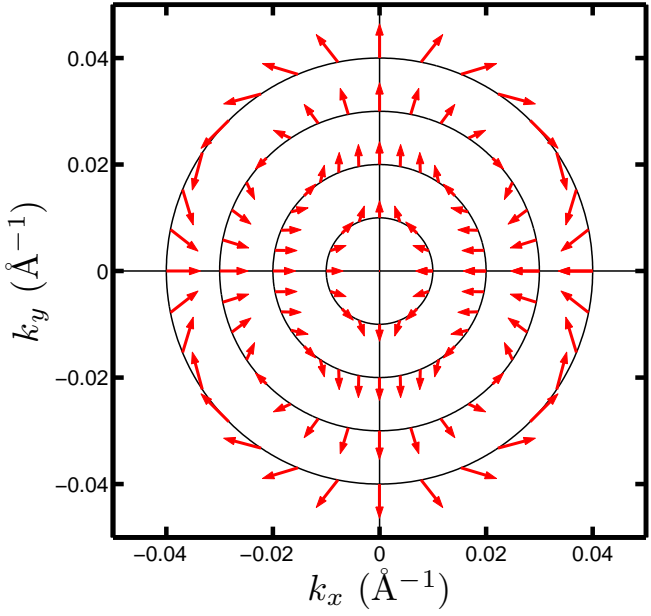} \caption{(Color online) {}Effective
magnetic field of a 100-$\operatorname{\text{\AA}}$-thick [001] InGaAs quantum
well with $\langle k_{z}^{2}\rangle=3.6\times10^{-4}$ $\operatorname{}^{-2}$.}%
\label{fig2}%
\end{figure}

For 2DES grown along [001] with Dresselhaus terms up to the $k^{3}$, Eq.
(\ref{Larmor [001]}) describes the effective magnetic field shown as Fig.
\ref{fig2}, which simulates a 100-$%
\operatorname{\text{\AA}}%
$-thick InGaAs quantum well with $\langle k_{z}^{2}\rangle=3.6\times10^{-4}$ $%
\operatorname{\text{\AA}}%
^{-2}$ (Ref. \onlinecite{Winkler03}). The CISP direction is opposite to the
effective magnetic field. Note that in Fig. \ref{fig2}, the field distribution
near the central region (small $k$) is dominated by the linear term
(\ref{Larmor D}) (cf. the right inset of Fig. \ref{fig3}).

In the rest of this paper, we focus on the Rashba and linear Dresselhaus [001]
terms. For effects with full SO terms in the Rashba-Dresselhaus systems, see
Refs. \onlinecite{Winkler03} and \onlinecite{Marques05}. The composite Larmor
frequency vector can be obtained by adding Eq. (\ref{Larmor D}) with $\hat
{n}=(0,0,1)$ and Eq. (\ref{Larmor R}) together,%
\begin{align}
\vec{\Omega}_{RD}^{[001]}  &  =\vec{\Omega}_{R}(\hat{n}=\hat{z})+\vec{\Omega
}_{D}^{[001]}\nonumber\\
&  =\frac{2}{\hbar}\left[  \alpha\left(  k_{y},-k_{x},0\right)  +\beta\left(
-k_{x},k_{y},0\right)  \right]  . \label{Larmor RD}%
\end{align}
The spin splitting linear in $k$ takes the form $\Delta_{\mathbf{k}%
}=\left\vert \zeta\right\vert k$ with $\zeta=i\alpha e^{-i\phi}+\beta
e^{i\phi}$. Thus the CISP in linear Rashba-Dresselhaus [001] 2DESs is
explicitly given by%
\begin{equation}
\left[  \mathbf{S}\right]  _{\mathbf{k}}^{RD001}=-\frac{\hbar}{2}\tanh
\frac{\left\vert \zeta\right\vert k}{k_{B}T}\hat{\Omega}_{RD}^{[001]}.
\label{Sk RD001}%
\end{equation}

\subsection{Remark on effective mass}

In general, the inplane effective mass $m$ of the electrons is not constant
but depend strongly on $\mathbf{k}$ for realistic semiconductor systems.
However, in the long-wavelength limit $k_{F}a\ll1$ ($k_{F}$ and $a$ the Fermi
wave vector and lattice constant, respectively), the effective mass, defined
by the inverse of the second derivative of $E(\mathbf{k})/\hbar^{2}$ with
respect to $k$, is a constant due to the parabolic nature of $E(\mathbf{k})$
solved from Hamiltonian (\ref{H}). In this limit, even though the band
structure can be anisotropic due to the interplay between different SO
couplings (such as Rashba plus linear Dresselhaus [001]), the effective mass
remains constant. In the present analysis, we work in this $k_{F}a\ll1$ limit,
within which the Hamiltonian (\ref{H}) is valid. Interestingly, our CEA
formulas such as Eq. (\ref{Sk 2}) do not contain the dependence of $m$.

Away from $k_{F}a\ll1$ region, the energy dispersion is no longer parabolic,
and the free-electron-like model Hamiltonian (\ref{H}) and hence the follow-up
derivations fail. Analysis of the CISP phenomenon requires other formalisms
such as the LKF, to be employed in the coming section. Nevertheless, we will
not look further into the influence of the $\mathbf{k}$-dependent effective
mass on the CISP.

\section{Numerical results: Landauer-Keldysh formalism\label{sec numerical}}

To inspect the validity of the previously proposed statistical picture and
further examine the CISP, we now perform local spin-density calculation in
finite-size 2DESs attached to four normal metal leads by using the
LKF.\cite{Datta95,Nikolic05,Nikolic06} \begin{figure}[t]
\centering \includegraphics[width=8.6cm]{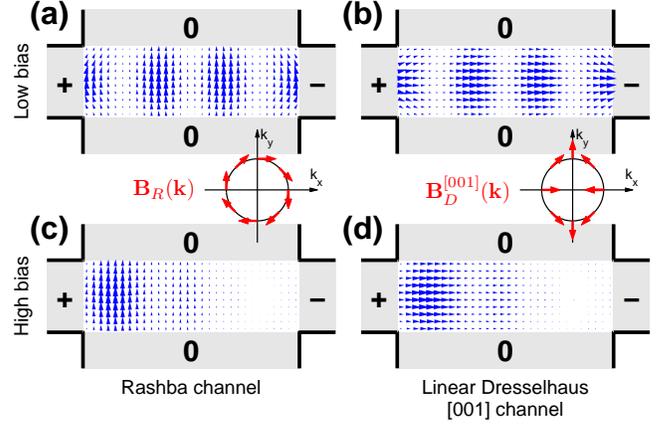} \caption{{}(Color online)
Spin orientation in a $30a\times10a$ channel with $a=1$ $\operatorname{nm}$.
Channels with linear Rashba model are considered in (a) and (c) while those
with linear Dresselhaus [001] model are in (b) and (d). The direction of each
sharp triangle represents the inplane spin vector $\langle\mathbf{S}%
\rangle_{\parallel}=(\langle S_{x}\rangle,\langle S_{y}\rangle)$ of the local
spin density. The size of the triangle depicts the magnitude of $\langle
\mathbf{S}\rangle_{\parallel}$. Effective magnetic fields due to individually
the Rashba and the Dresselhaus [001] fields are shown in the insets.}%
\label{fig3}%
\end{figure}

\subsection{Local spin densities in extreme Rashba and Dresselhaus [001]
cases}

As a preliminary demonstration, Fig. \ref{fig3} shows the position-dependent
in-plane spin vectors $\langle\mathbf{S}\rangle_{\parallel}^{\mathbf{r}%
}=(\langle S_{x}\rangle_{\mathbf{r}},\langle S_{y}\rangle_{\mathbf{r}})$, with
the local spin densities $\langle S_{x}\rangle_{\mathbf{r}}$ and $\langle
S_{y}\rangle_{\mathbf{r}}$ calculated by the LKF. Here we adopt the finite
difference method and discretize the $30a\times10a$ channel, made of
InGaAs/InAlAs heterostructure \cite{Nitta97} grown along [001], into a square
lattice with lattice spacing $a=1%
\operatorname{nm}%
$. Accordingly, this gives the kinetic and Rashba hopping strengths
$t_{0}\equiv\hbar^{2}/2ma^{2}=0.762$ $%
\operatorname{eV}%
$ and $t_{R}\equiv\alpha/2a=3.6$ m$%
\operatorname{eV}%
$, respectively. For the Dresselhaus SO coupling, we again assume the quantum
well thickness $d=100$ $%
\operatorname{\text{\AA}}%
$ and $\langle k_{z}^{2}\rangle\approx(\pi/d)^{2}$, and use $\gamma\approx27$
$%
\operatorname{eV}%
\operatorname{\text{\AA}}%
^{3}$ to give [see Eq. (\ref{beta})] $\beta=\gamma\left\langle k_{z}%
^{2}\right\rangle \approx2.\,\allowbreak66\times10^{-2}%
\operatorname{eV}%
\operatorname{\text{\AA}}%
$, resulting in the Dresselhaus hopping strength $t_{D}\equiv\beta/2a=1.33$ m$%
\operatorname{eV}%
$.

Let us first consider the extreme cases, pure Rashba and pure Dresselhaus
[001] channels. As expected, the spin vectors are mostly oriented antiparallel
to $\mathbf{B}_{\text{eff}}(\mathbf{k})$, which is, for $\mathbf{k}%
\parallel\hat{x}$, pointing to $-\hat{y}$ in the Rashba channel [Fig.
\ref{fig3}(a)/(c) with low/high bias], and $-\hat{x}$ in the Dresselhaus [001]
channel [Figs. \ref{fig3}(b)/(d) with low/high bias]. Here (and hereafter) the
low and high biases mean $eV_{0}=2$ m$%
\operatorname{eV}%
$ and $0.2$ $%
\operatorname{eV}%
$, respectively, and we label the applied potential energy of $\pm eV_{0}/2$
as \textquotedblleft$\pm$\textquotedblright,\ and $eV_{0}=0$ as
\textquotedblleft$0$\textquotedblright\ on each lead. Note that the spin
distribution, modulated by the charge distribution, forms standing waves in
the low bias regime since the electrons behaves quantum mechanically, while
that in the high bias regime, i.e., the nonequilibrium transport regime,
decays with distance.\cite{Liu07a} The polarization in the latter (high bias)
is about two orders of magnitude stronger than the former (low
bias).\begin{figure}[t]
\centering \includegraphics[width=8.6cm]{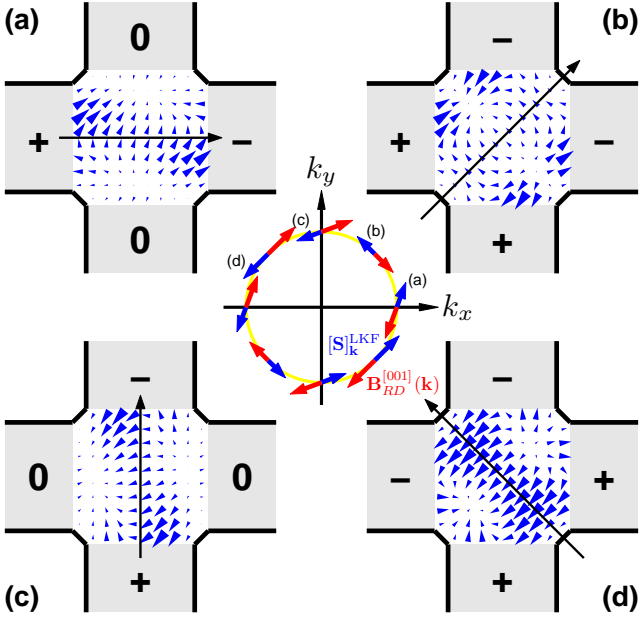} \caption{{}(Color online)
Local spin densities by LKF in a square Rashba-Dresselhaus [001] channel with
(a) left to right, (b) left-bottom to right-top, (c) bottom to top, and (d)
right-bottom to left-top bias configurations. Bias regime belongs to low:
$eV_{0}=2$ m$\operatorname{eV}$. Inset: $\left[  \mathbf{S}\right]
_{\mathbf{k}}^{\text{LKF}}$ vs. $\mathbf{B}_{RD}^{[001]}(\mathbf{k})$ in the
$k_{x}$-$k_{y}$ coordinate.}%
\label{fig4}%
\end{figure}

\subsection{Consistency check: Analytical canonical ensemble average vs
numerical Landauer-Keldysh formalism\label{consistency check}}

We now consider a four-lead square channel with coexisting Rashba and linear
Dresselhaus [001] terms. The coupling constants are set identical to those
introduced previously. Removing the four corner sites to avoid short circuit,
the sample size is $(10\times10-4)a^{2}$. To see if the CISP direction follows
the opposite effective magnetic field for all $\mathbf{k}$ directions, we
change the current direction by applying different bias configurations. As
shown in Figs. \ref{fig4}(a), (b), (c), and (d), the electrons flow from left
to right, from left bottom to right top, from bottom to top, and from right
bottom to left top, respectively. Other current directions are done in a
similar way, but not explicitly shown here. In averaging the in-plane local
spin densities $\langle S_{x}\rangle_{\mathbf{r}}$ and $\langle S_{y}%
\rangle_{\mathbf{r}}$ over all the lattice points at $\mathbf{r}$ within the
conducting sample, we compare in the inset of Fig. \ref{fig4} $\left[
\mathbf{S}\right]  _{\mathbf{k}}^{\text{LKF}}\equiv(\overline{\langle
S_{x}\rangle},\overline{\langle S_{y}\rangle})$ with the effective magnetic
field $\mathbf{B}_{RD}^{[001]}(\mathbf{k})=(\hbar/2\mu_{B})\vec{\Omega}%
_{RD}^{[001]}(\mathbf{k})$, where $\vec{\Omega}_{RD}^{[001]}(\mathbf{k})$ is
given by Eq. (\ref{Larmor RD}). As expected by our statistical picture
introduced in Sec. \ref{sec CEA}, $\left[  \mathbf{S}\right]  _{\mathbf{k}%
}^{\text{LKF}}$ arrows are all opposite to $\vec{\Omega}_{RD}^{[001]}%
(\mathbf{k})$ for all $\mathbf{k}$ directions, despite some indistinguishably
tiny differences. Note that the additive and destructive effects between the
two SO terms are also observed at $\pm$[\={1}10] and $\pm$[110], respectively.
Along $\pm$[\={1}10] ($\pm$[110]), strongest (weakest) spin splitting
$\Delta_{\mathbf{k}}$, and hence the CISP magnitude [Eq. (\ref{Sk 2})], occur.
Note that here we apply low bias. With high bias the results also agree
perfectly with the CEA picture (not shown).

\subsection{Bias dependence of current-induced spin polarization}

\begin{figure}[t]
\centering \includegraphics[width=8.5cm]{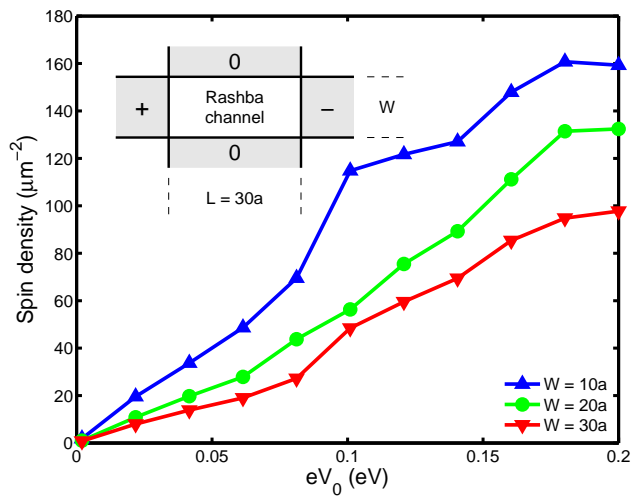} \caption{{}(Color online)
Bias dependence of spin densities induced by the CISP in Rashba 2DESs.}%
\label{fig5}%
\end{figure}

Having shown that the statistical argument indeed works well, we next examine
the bias dependence of the CISP, which is expected to be a proportional
relation, as has been experimentally observed.\cite{CISPexp1} We return to
Rashba channels. Spin densities, i.e., the total spin divided by the total
area of the conducting channel, obtained via $\sum_{\mathbf{r}}\langle
S_{y}\rangle_{\mathbf{r}}/(Na^{2})$ here with $N$ being the number of total
lattice sites in the conducting sample, are reported in Fig. \ref{fig5} for
sample widths $W=10a,20a,30a$. Sample length is set $L=30a$. Consistent to the
experiment, the calculated spin densities increase with $eV_{0}$. In addition,
linear response within $eV_{0}\lesssim0.1t_{0}=0.076$ $%
\operatorname{eV}%
$ is clearly observed in all cases. Nonlinearity enters when $eV_{0}$ grows so
that nonequilibrium statistics dominates. Note that the calculated local spin
density distribution satisfies the usual SHE symmetry,\cite{Nikolic06} so that
we have $\sum_{\mathbf{r}}\langle S_{x}\rangle_{\mathbf{r}}=\sum_{\mathbf{r}%
}\langle S_{z}\rangle_{\mathbf{r}}=0$ and $|\sum_{\mathbf{r}}\langle
\mathbf{S}\rangle_{\mathbf{r}}|=\sum_{\mathbf{r}}\langle S_{y}\rangle
_{\mathbf{r}}$.

\subsection{Comparison of current-induced spin polarization in semiconductor
heterostructures and metal/semimetal surface states}

Next we extend the calculation of the spin density due to the CISP to other
materials. In addition to semiconductor heterostructures, 2DESs have been
shown to exist also on metal surfaces supported by the surface
states.\cite{Davison92} Due to the loss of inversion symmetry, the metallic
surfaces may exhibit Rashba spin splitting as
well.\cite{Lashell96,Bihlmayer06} Here we consider three samples: $54\times18$
$%
\operatorname{nm}%
^{2}$ InGaAs/InAlAs heterostructure, $14.7\times4.9$ $%
\operatorname{nm}%
^{2}$ Au(111) surface, and $16.2\times5.4$ $%
\operatorname{nm}%
^{2}$ Bi(111) surface. We arrange the lead configuration of all the three
samples as those in Fig. \ref{fig3} and apply high bias. The sizes we choose
here are to maintain roughly the same lattice site number $N\lesssim1000$ and
keep the length-width ratio $\sim3$. Note that realistic lattice structure are
considered for the surface states [hexagonal for Au(111) and honeycomb for
Bi(111) bilayer], while finite-difference method based on the long-wavelength
limit for the heterostructure is adopted. For introductory reviews of those
surfaces, see Ref. \onlinecite{Reinert03} for noble metal surfaces, including
gold, and Ref. \onlinecite{Hofmann06} for bismuth surfaces.
\begin{ruledtabular}\begin{table}[<b>]%
%

\begin{tabular}
[c]{llll}%
Material & $\underset{\text{(heterostructure)}}{\text{InGaAs/InAlAs}}$ &
$\underset{\text{(surface state)}}{\text{Au(111)}}$ & $\underset
{\text{(surface state)}}{\text{Bi(111)}}$\\\hline
$m/m_{0}$ & $0.050$ & $0.251$ & $0.340$\\
$\alpha$ ($%
\operatorname{eV}%
\operatorname{\text{\AA}}%
$) & $0.072$ & $0.356$ & $0.829$\\
$E_{F}-E_{b}$ ($%
\operatorname{eV}%
$) & $0.108\,$ & $0.417$ & $0.083$\\
Reference &
\onlinecite{Nitta97}%
&
\onlinecite{Lashell96}%
&
\onlinecite{Koroteev04}%
\\
CISP ($10^{-3}%
\operatorname{nm}%
^{-2}$) & $0.240$ & $2.742$ & $8.382$%
\end{tabular}
\caption{Summary of effective mass ratio $m/m_{0},$ Rashba constant $\alpha
,$ Fermi energy $E_{F}$ (relative to the band bottom $E_{b}$), and the
calculated spin density due to CISP, for a set of materials.}\label{Table}%
\end{table}\end{ruledtabular}%

Band parameters extracted from experiments and the spin densities calculated
by the LKF are summarized in Table \ref{Table}. Clearly, the CISP increases
with the Rashba parameter $\alpha$. This suggests that the CISP (and actually
also the SHE) should be more observable on these surfaces. The recently
discovered Bi/Ag(111) surface alloy that exhibits a giant spin
splitting\cite{Ast07} is even more promising, but we do not perform
calculation for this interesting material here.

\subsection{Application of current-induced spin polarization: Generation of
in-plane spin-Hall pattern\label{CISP application}}

Finally, we propose an experimental setup, as an application of the CISP, to
generate an antisymmetric edge spin accumulation in the inplane component,
i.e., an inplane spin-Hall pattern. For simplicity, let us consider a Rashba
2DES with the parameters for the LKF calculation taken the same as those in
Fig. \ref{fig5}. Sample size is about $30\times30$ $%
\operatorname{nm}%
^{2}$. We apply high bias of $eV_{0}=0.2$ $%
\operatorname{eV}%
$ and arrange a special bias configuration.\begin{figure}[t]
\centering \includegraphics[width=8.65cm]{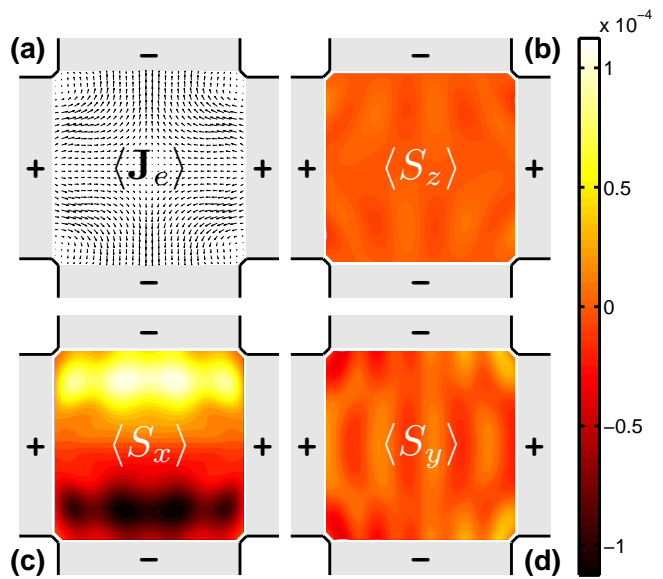} \caption{{}(Color online)
Mapping of the (a) local charge current density, local spin densities (b)
$\langle S_{z}\rangle,$ (c) $\langle S_{x}\rangle,$ and (d) $\langle
S_{y}\rangle$ in a four-terminal square channel with a special bias
arrangement. Unit in (b)--(c) is $\hbar/2$.}%
\label{fig6}%
\end{figure}

As shown in Fig. \ref{fig6}(a), unpolarized electron currents are injected
from the left and right leads and are guided to the top and bottom ones. Under
such design, the spin accumulation in $\langle S_{z}\rangle_{\mathbf{r}}$
exhibits merely a vague pattern [see Fig. \ref{fig6}(b)]. Contrarily, the
pattern of $\langle S_{x}\rangle_{\mathbf{r}}$ shows not only antisymmetric
edge accumulation in the channel but also magnitude much stronger than the
out-of-plane component [see Fig. \ref{fig6}(c)]. This pattern is reasonably
expected by the CISP due to the opposite charge flows along $\pm\hat{y}$ at
the top and bottom edges, and hence resembles an inplane SHE.

In determining $\langle S_{y}\rangle_{\mathbf{r}}$, Fig. \ref{fig6}(d) does
not show a rotated pattern from $\langle S_{x}\rangle_{\mathbf{r}}$ due to the
nonequilibrium transport. In the nonequilibrium transport regime, a distance
apart from the source leads is required to induce the CISP, and therefore no
significant $\langle S_{y}\rangle_{\mathbf{r}}$ is observed near the source
(left and right) leads. This can be seen by comparing the local spin density
distributions in the low-bias and high-bias regimes shown in Figs.
\ref{fig3}(a) and \ref{fig3}(b), and Figs. \ref{fig3}(c) and \ref{fig3}(d), respectively.

\section{Summary\label{sec summary}}

In conclusion, we have rederived the CISP due to SO coupling in the absence of
impurity scattering based on equilibrium statistical mechanics.
Correspondingly, a simple picture (Fig. \ref{fig1}) valid for both bulk
structures and 2DESs is proposed to help qualitatively explain the CISP. Our
explanation for the spin polarization of the moving electron ensemble in solid
due to effective magnetic field is an exact analog to that of the rest
electron ensemble in vacuum due to external magnetic field.\cite{Sakurai94}
The picture is further tested to work well\ even in the regime of
nonequilibrium transport in finite-size samples, by employing the numerical
LKF. Extending the spin density calculation from the semiconductor
heterostructure to metal and semimetal surface states, our calculation
confirms that the polarization increases with the SO coupling strength, and
hence suggests that the CISP should be more observable on metal and semimetal
surfaces with stronger Rashba SO coupling.\cite{Lashell96,Koroteev04,Ast07} As
an application of the CISP, we also suggest an interesting bias configuration
for the four-terminal setup to generate inplane SHE [Fig. \ref{fig6}(c)].

\begin{acknowledgments}
One of the authors (M.H.L.) appreciates S. D. Ganichev and L. E. Golub for
stimulating discussion, and L. Ding and G. Bihlmayer for useful information.
Financial support of the Republic of China National Science Council Grant No.
95-2112-M-002-044-MY3 is gratefully acknowledged.
\end{acknowledgments}

\bibliographystyle{apsrev}
\bibliography{mhl2}

\end{document}